\documentclass[%
 reprint,
 amsmath,amssymb,
 aps,
prb,
]{revtex4-1}

\usepackage{graphicx}
\usepackage{dcolumn}
\usepackage{bm}

\def\NpPtGa{Np$_2$PtGa$_3$}
\def\NpPdGa{Np$_2$PdGa$_3$}
\def\UPdGa{U$_2$PdGa$_3$}

\begin{document}

\preprint{APS/123-QED}

\title{Ferromagnetic behavior of the Kondo lattice compound \NpPtGa}

\author{V. H. Tran,$^1$ J. -C. Griveau,$^2$ R. Eloirdi,$^2$ and E. Colineau$^2$ }
\address{$^1$Institute of Low Temperature and Structure Research, Polish Academy of Sciences, P. O. Box 1410, 50-950 Wroc\l aw, Poland\\
$^2$ European Commission, Joint Research Centre, Institute for Transuranium Elements, Postfach
2340, D-76125 Karlsruhe, Germany\\
}





\begin{abstract}
In this study we report the results of study of novel ternary \NpPtGa~compound. The x-ray-powder diffraction analysis reveals that the compound crystallizes in the orthorhombic CeCu$_2$-type crystal structure (space group Imma) with  lattice parameters \emph{a} = 0.4409(2) nm, \emph{b} = 0.7077(3) nm and \emph{c }= 0.7683(3)  nm at room temperature. The measurements of dc magnetization, specific heat and electron transport properties in the temperature range 1.7 - 300 K and in magnetic fields up to 9 T imply that this intermetallic compound belongs to a class of ferromagnetic Kondo systems. The Curie temperature of $T_C \sim$ 26 K is determined from the magnetization and specific heat data.  An enhanced coefficient of the electronic specific heat of $\gamma$ = 180 mJ/(mol at. Np K$^2$) and -ln\emph{T} dependence of the electrical resistivity indicate the presence of Kondo effect, which can be described in terms of the S = 1 underscreened Kondo-lattice model. The estimated Kondo temperature $T_K \sim$ 24 K,  Hall mobility of $\sim$ 16.8 cm$^2$/Vs and effective mass of $\sim$ 83 $m_e$ are consistent with assumption that the heavy-fermion state develops in \NpPtGa~at low temperatures. We compare the observed properties of \NpPtGa~to that found in \NpPdGa~and discuss their difference in regard to change in the exchange interaction between the conduction and localized 5f electrons. We have used the Fermi wave vector $k_F$ to evaluate the Rudermann-Kittel-Kasuya-Yosida (RKKY) exchange. Based on experimental data of the (U, Np)$_2$(Pd, Pt)Ga$_3$ compounds we suggest that the evolution of the magnetic ground states in these actinide compounds can be explained within the RKKY formalism.
\begin{description}

\item[PACS number(s)]
 71.27.+a, 75.20.Hr, 75.30.-m
\end{description}
\end{abstract}

\maketitle


\section{Introduction}
\par f-electron intermetallics exhibit a variety of exotic phenomena including non-Fermi liquid behavior, heavy-fermion, unconventional superconductivity, etc.. Nevertheless, their magnetic nature remains still unresolved issue due to the complexity of different interactions and the interplay with Kondo effect. In order to characterize complex phenomena, in parallel with theoretical research, experimentally comparative study of physical properties of the intermetallics can often be undertaken. Exemplary investigations are just these on various series of 5f-electron intermetallics (see [1]). One of the most important  conclusions is that 5f-ligand hybridization is the principal  mechanism determining the magnetic order,  and  it causes the compounds with light actinides (U,  Np) to have a narrow 5f  band at the Fermi level.
Furthermore, the 5f electrons in Np-based compounds have usually less itinerant character than those in
the corresponding U ones. Results of our systematic studies on  213-type stoichiometric compounds U$_2$TGa$_3$ (T = d-electron transition metals),[2] U$_2$(Pd, Pt)Ga$_3$,[3] and \NpPdGa,[4]  have
indicated that the degree of the 5f localization follows the general trend in actinide intermetallics mentioned  above.  In the recent work [4] we have discovered that  \NpPdGa~ is a ferromagnetic Kondo lattice (FKL). The occurrence of a ferromagnetic ground state in this compound was explained with help of an oscillating character of the Rudermann-Kittel-Kasuya-Yosida  (RKKY) interaction.  What more renders \NpPdGa~an interesting intermetallic compound is i) a competition of, at least, three  different  types of interactions -  Kondo effect, RKKY and crystal electric field (CEF), represented by  an energy relationship $T_K <  T_{RKKY} \sim \Delta_{CEF}$, ii) ferromagnetic Kondo lattice with considerably enhanced effective mass of the charge carriers and iii) atomic disorder enabling to form random magnetic interactions. In view of the difference in the d-electron numbers and covalent radii between Pd and Pt atoms involved in compounds, it is of interest to study a compound \NpPtGa, aiming to make a comparative investigation of the magnetic and electronic properties of these Np-based 213-type intermetallics. It was the original purpose of this paper to report the synthesis and crystallographic characterization, as well as to present the results of magnetic, specific heat and electron  transport measurements of a novel \NpPtGa~compound.
\par Our measurements of bulk properties (see below) show that \NpPtGa~orders ferromagnetically at 26 K. Additionally, the coefficient of the electronic specific heat of $\gamma$ = 180 mJ/(mol at.Np K$^2$) is sizeably enhanced, together with -ln\emph{T} dependence of the electrical resistivity, classifying \NpPtGa~to belong to relatively rare family of ferromagnetic Kondo lattices. It is worthwhile to mention that the coexistence of ferromagnetic order and Kondo effect is most frequently found in Ce-based, [5-11] or U-based systems, [12-16] but much rarer in Np-compounds, since only two compounds NpNiSi$_2$,[17] and Np$_2$PdGa$_3$,[4] were reported so far. Moreover, it seems that there should be different approaches to ferromagnetic Kondo problems as regards to particular values of spin angular momentum S of systems. For S = 1/2 systems of the Ce-based compounds, both ferromagnetic order and Kondo effect are governed by the s - f  exchange interaction between the localized S  = 1/2 spins and the conduction electrons.  Due to the competition of magnetism and Kondo effect, the magnetic ordering of the S = 1/2 systems appears at relatively low temperatures (less than 10 K), and the systems can be nonmagnetic if the localized moments are completely screened by the conduction electrons.[18]  Otherwise, in the U- and Np-based compounds with 5$f^n$, \emph{n} = 2 or 3 electrons, these larger spins (S $\geq$ 1) are partially screened by the conduction electrons only.  Appropriately, it is intuitively expected that the coexistence of the Kondo effect and ferromagnetic order may yield a high Curie temperature. In fact, hitherto $T_C$  in the ferromagnetic U and Np-based Kondo lattices was reported to attain values between 50 and 100 K, i.e., of one order of magnitude larger than those in the Ce-based compounds. In order to interpret the Kondo effect in the actinide ferromagnets, Coqblin  and coworkers. developed so-called underscreened Kondo lattice  (UKL) model.[19, 20]  Recently, the authors have obtained results for the case, where the Kondo temperature is smaller than the Curie temperature, and have compared to several bulk properties of NpNiSi$_2$, like magnetization, electrical resistivity and specific heat.[21]  Therefore, it is very encouraging to study \NpPtGa, not only as the physical properties can be systematically investigated in the isostructural 213-type stoichiometric group, but also \NpPtGa~as a new member of FKL and it serves to test theoretical models.

\section{Sample preparation and characterization}
\par A polycrystalline  sample  of  \NpPtGa~(0.5 g)  was  synthesized  from  high-purity
elements Np 3N, Pt   4N,  and  Ga  6N  by  arc-melting  under  Ti-gettered argon  purified atmosphere.  To  improve  homogeneity,  the sample was turned over and remelted several times. The ingot was wrapped in Ta foil and was annealed at 800 $^\circ$C for 14 days in evacuated quartz tube. The weight loss during the preparation was less than 0.5 $\%$. The quality of the sample was characterized by standard x-ray powder diffraction. The analysis of the x-ray diffraction pattern revealed that the majority phase ($>$  95  mass.  \%)  has  the  orthorhombic  CeCu$_2$-type  structure,  isostructural with U$_2$(Pd,Pt)Ga$_3$,[3] and \NpPdGa.[4] Refinement of crystal structure  of \NpPtGa~based on 63 observed Bragg reflections was done with the Rietveld method using Fullprof programm.[22] The Bragg reflections were fitted with Pseudo-Voigt function profile and the final refinements converged at R-factor = 1.016 and RF-factor = 0.6762.  We obtained lattice parameters \emph{a} = 0.4409(2) nm, \emph{b} = 0.7077(2) nm and \emph{c} = 0.7683(3)  nm. A comparison to the lattice parameters of \NpPdGa~(\emph{a} = 0.4445, \emph{b} = 0.7089 and \emph{c} = 0.7691 nm,[4]) discloses that the unit cell volume of the studied compound shrinks by 1 $\%$ relative to that of  \NpPdGa. Impurity phases present in our sample were identified to be NpO$_2$ (\emph{a} = 0.5433 nm, cubic structure with Fm-3m space group) and NpPtGa (\emph{a} = \emph{b} = 0.6733, \emph{c} = 0.3837 nm, hexagonal ZrNiAl-type structure with space group P6-2m). In the literature, NpO$_2$ was reported to undergo into an electric quadrupole state below $T_0$ =25  K,[23] whereas no information about the magnetic ground state of NpPtGa is available.

\section{Physical properties}
\subsection{Magnetic susceptibility and magnetization}
\par Dc magnetization (\emph{M}) measurements were carried out in the temperature range 1.72 - 300 K and in magnetic fields up to $\mu_0H$ = 7 T, using a commercial Quantum Design SQUID magnetometer. The \NpPtGa~sample was encapsulated in a specially designed sample holder. The response of the holder for magnetization was measured separately before inserting the sample and then subtracted from measured raw data. It turned out that the magnetic susceptibility, defined as $\chi = M/H$, of \NpPtGa~above 80 K is independent of applied magnetic fields and has a value of 3.26 $\times$ 10$^{-3}$ emu/(mol at. Np) at room temperature. Fig. 1 shows the temperature dependence of the inverse magnetic susceptibility obtained in a field of 7 T. The high-temperature data were fitted with the modified Curie-Weiss law $\chi(T) = \chi_0 + N_A\mu_{eff}^2/[3k_B(T-\theta_p)]$, where $N_A$ is the Avogadro number and $k_B$ is the Boltzmann constant. The fit yields the temperature independent susceptibility $\chi_0$ = 1.04(0.05)$\times$10$^{-3}$ emu/(mol at. Np), effective moment $\mu_{eff}$ = 2.28 $\pm$ 0.1 $\mu_B$/Np and paramagnetic Curie temperature $\theta_p$ = 33 $\pm$ 1 K. The temperature independent term $\chi_0$ usually accounts for the Pauli-type and Van Vleck paramagnetic susceptibilities. In the case of \NpPtGa, a large value of $\chi_0$ points to significant contribution of the conduction electrons to the density of states. After subtraction of $\chi_0$, the susceptibility data of \NpPtGa~(triangle symbols in Fig. 1) exhibit the Curie-Weiss behavior, indicative of the localized character of the 5f electrons. However, in comparison with the theoretical value of 2.68 $\mu_B$/ Np for the electronic configuration 5f$^4$ in the Russell-Saunders coupling, the observed $\mu_{eff}$ value is noticeably reduced. This means that the 5f electrons are partly delocalized, presumably due to a strongly electron correlation effect, such as the hybridization of the 5f- with the conduction electrons and/or Kondo screening by the conduction electrons. The observed large and positive paramagnetic Curie temperature indicates a dominant ferromagnetic exchange interaction between the magnetic Np moments in \NpPtGa. This interaction becomes stronger with decreasing temperature and finally causes an anomaly at low temperatures. As shown in the inset of Fig.  1, magnetization measurements at $\mu_0H$ = 0.02 T under both zero-field-cooled (ZFC) and field-cooled (FC) conditions yield a broad maximum at $T_{max}$ = 24 K. Above $T_{max}$ the magnetization curves have the same temperature dependencies, but below this temperature the curves separate from each other leading to the occurrence of irreversible magnetization.

\par To gain an information about the magnetic ground state, measurements of iso-field magnetization for selected fields in the temperature range 1.72 - 60 K as well as isothermal magnetization for several temperatures in fields up to 7 T were carried out. The left panel of Fig. 2 shows the experimental iso-field data. It is observed that the maximum of \emph{M(T)}  broadens upon application of higher field strengths and its precise position becomes hardly determined for fields larger than 5 T. In such large magnetic fields no more maximum occurs in the $M_{FC}(T)$-curves. Instead, a plateau forms in the low temperature range. It is worthwhile to distinguish two characteristic temperatures $T_{ir}$ and $T_{inf}$, which are obviously dependent on the applied fields. $T_{ir}$ symbolizes the temperature where the irreversibility of magnetization sets in, whereas $T_{inf}$ is defined as the minimum of the temperature derivative of the magnetization. As shown in the right panel of Fig. 2, $T_{ir}$ shifts down to lower temperatures, while $T_{inf}$ moves upward to higher temperatures. These features resemble the behavior of spin glasses,[24,25]  ferromagnets with a large magnetocrystalline anisotropy,[26] or with a random magnetic anisotropy, canted ferromagnets or antiferromagnets with competing antiferromagnetic and ferromagnetic interactions.[27,28]

\par Figure 3 shows isothermal magnetization data at several temperatures up to 50 K. Concerning the 2 K-data, there is no saturation even in a high magnetic field of 7 T. The tendency towards saturation of the magnetization suggests to invoke the local and magnetocrystalline anisotropy in the estimation of the saturated moment $M_s$. From fitting of the 2 K-magnetization data to the formula: [29]
\begin{equation}
M(H) = M_s + a/H + b/H^2,
\label{(1)}
\end{equation}
where the terms $a/H$ and $b/H^2$ correspond to the local and magnetocrystalline anisotropy, respectively, we obtain  $M_s$ = 0.9 $\mu_B$/Np. This value is considerably reduced if compared to the theoretical value 2.4 $\mu_B$ for Np$^{3+}$. The finding may imply an influence of the Kondo-type screening of the conduction electrons on the localized magnetic Np moments.

\par The magnetization data obtained below 10 K presented in Fig. 3 a remark a sharp kink at a critical field $H_{cr}$, resembling the behavior of the metamagnetic transition. The simplest assumption to reconcile this metamagnetic-like transition is the spin flip transition in an antiferromagnetic structure formed by magnetic Np moments. The other way to understand the magnetization behavior is to assume the effect of magnetocrystalline anisotropy. Indeed, we determined values of $H_{cr}$ as the peak of the field derivative of the magnetization (see bottom panel of Fig. 3), which do not depend regularly of the applied magnetic fields. In fact, $H_{cr}$ attains its maximum value at \emph{T} = 5 K, pointing rather to domain-wall motion and different types of magnetic anisotropy, than due to a change in the spin configuration affected by magnetic fields. Moreover, the metamagnetic-like transition is observed only for increasing applied fields. During decreasing fields, there appears considerable hysteresis and remanent magnetization. This behavior is consistent with the irreversibility of magnetization observed below $T_{ir}$ giving support for  magnetocrystalline effect present in ferromagnets.
\par The magnetization measurement at 25 K (see \emph{dM/dH} in Fig. 3 b) indicates that the magnetic moment of Np$^{3+}$ does not vanish yet. Above this temperature, e.g., at 50 K the \emph{M(T)} curve shows neither metamagnetic-like transition, hysteresis nor remanence, that is typical for the paramagnetic state.

\par  In Fig. 4, we plot $M^2$ as a function of $H/M$, so-called Arrott plot, for selected data obtained during decreasing applied magnetic fields. It is observed that the slope of the $M^2( H/M)$ -curves is positive and the extrapolation of the 25 K-data yields the intercept close to zero $\mu_B$.  These features serve as a convincing evidence for a ferromagnetic order in \NpPtGa~with the Curie temperature of approximately $T_C \sim$ 25 K. The plot gives a low remanent magnetization at 2 K ( $\sim$ 0.37 $\mu_B/Np$ ), which would suggest a small value of the ordered moment.
\par It is worthwhile to recall that the magnetic behavior of NpO$_2$, which was identified as an impurity in the studied sample, is unable to account for the magnetic anomalies observed here. Since, NpO$_2$
exhibits multipolar order at 25~K, with a very small induced dipolar magnetic moment and antiferromagnetic-like signatures in bulk properties.[30-32]

\subsection{Specific heat}
\par Specific heat measurements were performed using the 2-$\tau$ thermal relaxation method. The experiments were conducted in the temperature range 1.8 - 300 K and in magnetic fields up to 9 T, utilizing a commercial Quantum Design PPMS platform. The experimental data at 0 T after correction for the addenda and sample holder are presented in Fig. 5. We have analyzed the high-temperature specific heat data, assuming the presence of two contributions: electronic and phononic. These contributions are described by the $\gamma_{HT}$  coefficient and  the high-temperature Debye temperature $\Theta_D^{HT}$, respectively. The lattice specific heat is expected to follow the Debye function:
\begin{equation}
C_{ph}(T)=9Rn(T/\Theta_D^{HT})^3\int_0^{\Theta_D^{HT}/T}\frac{x^4 {\rm exp}(x)}{[{\rm exp}(x)-1]^2}dx,
\label{2}
\end{equation}
 where \emph{R} is the gas constant and \emph{n} is the number of atoms per formula unit. A fitting the specific heat data to the sum of $C_{ph}(T)$ and $\gamma_{HT}T$ yields $\Theta_D^{HT}$ = 230 $\pm 1$ K and $\gamma_{HT}$ = 5 $\pm 0.5$ mJ/(mol at.Np K$^2$). The dashed line shown in Fig. 5 is the result of the fit.
\par The temperature dependence of $C_p(T)$  in a low-enough  temperature regime is usually expected to satisfy the standard relation $C_p(T) = \gamma T + \beta T^3$ for metals, where the first term is the contribution of the conduction electrons and the latter is from the phonons. However, the $C_p/T$ data of \NpPtGa~exhibit an upturn with decreasing temperature. It is seen also that the upward curvature of $C_p/T$ shifts to higher temperature when a magnetic field of 9 T is applied (see inset). Because the $^{237}$Np isotope with the nuclear moment $\mu_I$ =3.14 $\mu_N$, may have six nuclear energy levels ranging from \emph{I} = -5/2 to 5/2, one can invoke the splitting of the nuclear ground-state level of  the  $^{237}$Np nuclei to explain the low-temperature specific heat of \NpPtGa. Usually, the nuclear specific heat $C_N(T)$ can be modeled with the Schottky specific heat $C_{Sch}(T)$. In a simple two-level  system, $C_{Sch}$ is given by
\begin{equation}
C_{Sch} (T)= R(\frac{\varepsilon}{k_BT})^2(\frac{g_0}{g_1})\frac{{\rm{exp}}(\frac{\varepsilon}{k_BT})}{1+\frac{g_0}{g_1}{\rm{exp}}(\frac{\varepsilon}{k_BT})},
\label{(3)}
\end{equation}
where $g_0$ and $g_1$ are the degeneracies of lower and upper levels and  $\varepsilon/k_B$ is an energy separation between these levels. Taking into account the electronic, phononic and nuclear contributions, the low-temperature specific heat can be written under the form:
\begin{equation}
C_p(T)/T = \gamma + \beta T^2 + C_N/T.
\label{(4)}
\end{equation}

We fitted the specific heat data in the temperature range 1.8 - 5 K to Eq. [\ref{(4)}] (see the inset of Fig. 5). As a result of the fitting we get $\varepsilon/k_B$ = 0.37 K in 0 T and  $\varepsilon/k_B$ = 0.40 K in 9 T. The splitting energy is associated with the hyperfine field $B_{hf}$ via the relation:[33]  $\varepsilon = \frac{R}{3k_B}(\frac{I+1}{I}) (g_N\mu_I \mu_N B_{hf})^2$, where $g_N$ = 0.793 is the gyromagnetic factor of $^{237}$Np. From the experimental $\varepsilon$ values we inferred the effective magnetic field  $B_{hf}$ = 214 in zero applied field and 223 T in an applied field of 9 T, corresponding to an ordered magnetic moment of $\sim$ 1 $\mu_B$, if one assumes 1 $\mu_B$ = 215 T. As expected, the coefficient $\beta$ = 0.57 mJ/(mol at.Np K$^4$) is independent of the applied magnetic field. From the $\beta$ value and using the formula: $\Theta_D = [n12\pi^4\frac{R}{5\beta}]^{1/3}$, we calculated the low-temperature Debye temperature $\Theta_D$ = 217 K, which is in good agreement with that deduced from the high-temperature data. A remarkable result of our experiment is that the Sommerfeld coefficient $\gamma$ = 180 mJ/mol.Np K$^2$ is obtained from the fitting of the data in zero field, whereas $\gamma$ = 186 mJ/(mol at. Np K$^2$) from the data in 9 T. The sizeably enhanced Sommerfeld coefficient manifests a strong correlation of the Np - 5f  electrons and this implies that \NpPtGa~can be classified to the class of the Np-based heavy-fermion compounds, like e.g., NpPd$_5$Al$_2$ with $\gamma$ = 190 - 200 mJ/(mol K$^2$).[34]

\par A careful inspection of the experimental data shown in Fig. 6 reveals an anomalous behavior of \NpPtGa~associated with magnetic orders. We recognize two specific heat anomalies, at 26 K and 30 K, respectively. From the temperature derivative of the ratio $C_p(T)/T$ (see inset of the figure), we suspect that the first transition is of the second-order while the latter is of the first-order. Based on the magnetic data and analysis of the specific heat data (see below) we suggest that the anomaly at 26 K may be ascribed to a ferromagnetic order of the Np moments. The broadened and smearing transition is presumably due to disorder effects on magnetic interactions. Unfortunately, we have no explanation on the origin of the anomaly at 30 K, denoted as $T_i$. A comparison of the 0 T and 9 T data concludes that the application of a magnetic field of 9 T has no influence on the position of these anomalies. This fact could exclude spin-glass type as an origin of the anomaly at 26 K. Concerning the field dependencies of $T_{ir}$ and $T_{inf}$ of the \emph{M(T)} curves, two comments can be made. We cannot ascribe neither $T_{ir}(H)$ nor $T_{inf}(H)$ to field dependencies of the magnetic phase transition temperature $T_C$. In our opinion, the reasons of the irreversibility $T_{ir}$ may be magnetocrystalline anisotropy as we discussed earlier. On account of  $T_{inf}(H)$, we suspect it is due to the field-induced ferromagnetic short-range correlations.

\par In Fig. 7 we show the temperature dependence of the 5f-electron specific heat, which is obtained by the subtraction of the nuclear, phononic and high-temperature electronic contributions from the total specific heat. Clearly, the specific heat jump in the $C_{5f}(T)/T$ at 30 K is negligible and one reasonably ascribes to the contribution of an impurity. It appears that the $C_{5f}(T)/T$ exhibits a small jump at $T_C$ = 26 K, which may support a long-range magnetic order in the studied compound. However, bearing in mind eventual contribution of impurity NpO$_2$ phase possessing similar value of phase transition temperature, we should take the data with caution.  When comparing the experimental data with that reported for NpO$_2$,[36,37] we see that for \NpPtGa, the transition at $T_C$ is rather broad, in contrast to very sharp in NpO$_2$. Nonetheless, if NpO$_2$ contributes to the measured specific heat, then its content is of the order of 1 $\%$ of the investigated sample, since the specific jump at 26 K in our sample is $\Delta C_{mag}/T$ = 0.88$\times$10$^{-2}$ J/(mol K$^2$), compared to 0.86 J/(mol K$^2$) found for NpO$_2$ at 25.5 K.
 \par We now consider the implications of the $C_{5f}(T)/T$ data, which exhibit a broad hump at about $T_{max}$ = 30 K just above $T_C$ and a long tail above at higher temperatures. At first sight, it is likely that crystal electric field splitting and/or short-range magnetic interactions contribute to the specific heat. If we assumed a doublet-doublet as possible CEF scheme and taking the position of $T_{max}$, we would expect a CEF splitting energy $\Delta_{CEF}/k_B$ = 82 K. The CEF specific heat with such $\Delta_{CEF}/k_B$ is illustrated as dotted line in Fig. 7. However, as can be seen, the contribution of such a CEF splitting is hard to reproduce fully the specific heat tail. Therefore, it is possible that the 30 K - hump and heat capacity tail in \NpPtGa~are due to the short-range magnetic correlations. The specific heat enhancement due to short-range magnetic correlations was studied theoretically, among other researches, by  Bonner and Fisher.[38]
\par Below $T_C$, the 5f-electron specific heat consists, at least, of two components $C_{mag}$ and $C_K$. The first is due to the magnon excitations, being insignificant at temperatures far below $T_C$. Conversely, the second component is supposing Kondo effect, which confers the large value of the Sommerfeld ratio. This contribution is dominating at low temperatures but decreases with increasing temperature. Within in the Coqblin-Schrieffer model, the Kondo impurity problem has been numerically calculated for several angular momenta\emph{ J} by Rajan.[39] If we take the Kondo temperature $T_K$ = 24 K and ground state \emph{J } = 1/2 we may approximate the Sommerfeld ratio at low temperature, and then the Kondo specific heat $ C_K(T)$ in \NpPtGa, represented in Fig. 7 by the dashed line. Assuming that $C_{5f}(T) = C_K(T) + C_{mag}(T)$ we have extracted the magnetic specific heat $C_{mag}(T)$ depicted in Fig. 7 as closed circles. For data far below $T_C$,  we have fitted the $C_{mag}(T)$ data using the standard formula for the magnon specific heat with an anisotropy energy gap $\Delta$ in magnon spectrum :
\begin{equation}
C_{mag}(T) = \alpha T^{3/2}{\rm exp}(-\Delta/k_BT),
\label{(5)}
\end{equation}
where $\alpha$ is a constant. Fitting the experimental magnetic specific heat $C_{5f}(T)$ with Eq. [\ref{(5)}] gives $\alpha$ = 0.03 $\pm$ 0.006 J/(mol at. Np K$^{5/2}$) and $\Delta/k_B$ = 4.9$\pm$ 0.2 K.
\par The 5f-electron, magnetic and Kondo entropies are calculated by integrating $C_{5f}(T)/T$, $C_{mag}(T)/T$ and $C_K(T)/T$
with respect to temperature and are plotted in the inset of Fig. 7 as solid, dotted and dashed lines, respectively. At $T_C$, $S_{mag}(T)$ reaches 4 J/(mol K), i.e. about 70 $\%$ of  Rln2 - the expected value of the entropy gain from a doublet ground state. The reduced magnetic entropy can be attributed to the influence of the Kondo effect.

\subsection{Electron transport properties}
\par Measurements of electron transport properties were carried out by means of the standard ac four-probe technique using a Quantum Design PPMS platform. The measurements were performed from 1.8 K to 300 K and in magnetic fields from  -9 T to 9 T,  with an excitation current of \emph{j} = 0.5 mA applied perpendicularly to the direction of magnetic fields.  The temperature dependence of the electrical resistivity of \NpPtGa~is shown in Fig. 8. The resistivity at room temperature amounts to 264 $\mu \Omega$ cm. It increases logarithmically with decreasing temperature, in agreement with the single-site scattering of the Kondo effect. After showing a maximum at 14 K and minimum at 7 K, the resistivity increases logarithmically again with further decreasing temperature. Such behavior of the resistivity can be attributed to the Kondo-lattice formation and a competing influence of the Kondo and CEF effects. According to the theory developed by Cornut and Coqblin,[40] the scattering of the conduction electrons in the presence of well separated crystal-field levels, would affect the \emph{C}ln\emph{T} dependence of the resistivity in two temperature regions, where the coefficient \emph{C} is given by
\begin{equation}
C \propto J_{cf}^3\frac{\lambda_i^2-1}{2J+1}.
\label{(6)}
\end{equation}
$J_{cf}$ is the exchange interaction between the conduction and localized 5f electrons, $\lambda_i$ are the effective degeneracies of the crystal field 5f levels and \emph{J} is the total angular momentum of the Np$^{3+}$ ions. We calculated the ratio of slopes of the ln\emph{T}-dependencies at low temperature and at high temperature regions $\nu = (\lambda_l^2-1)/(\lambda_h^2-1)$  = 0.075, being close to that of a scheme where the 5f level splits in the doublet and the sextet states.

\par We wish to point out that the Curie temperature $T_C$ is difficult to determine from the electrical resistivity data. This does not rule out the possibility of magnetic phase transition at $T_C \sim$ 26 K. The two following factors can affect the lack of anomaly at $T_C$. The random distribution of Pt and Ga atoms and a tremendous contribution of the Kondo effect. As we will show below, the Kondo energy scale in this compound is comparable to the magnetic interaction strength. The inset of Fig. 8 compares the low-temperature resistivity of \NpPtGa~at 0 and 9 T. We observe that the applied magnetic field reduces the resistivity. The behavior is consistent with the ferromagnetic and Kondo properties of the compound.
\par Fig. 9 shows the magnetoresistance (\emph{MR}), which is defined as $MR = \frac{\rho(T, H) - \rho(T, 0)}{\rho(T, 0)}$.  As can be seen, \emph{MR} is negative in the range of fields 0 - 9 T. Below $T_C$, \emph{MR} reaches as large as - 3.6 $\%$ at 9 T. It is noted that the shape of the field dependence of the \emph{MR} suggests the electron scattering at the ferromagnetically coupled Np moments under fields. In the range of magnetic fields -9 - 0 T, the \emph{MR} data exhibit an irreversibility effect. Especially, the data obtained at 2.5 K display a positive maximum at -3.5 T. The position of this maximum moves to 0 T as temperature increases. The \emph{MR} behavior may be correlated with the magnetization reversal in the "metastable" ferromagnetic state, accompanied by the magnetization hysteresis. The \emph{MR} data at 50 K, i.e. above $T_C$ are due to short-range ferromagnetic correlations.
\par

\par  The Hall voltage $V_H(H)$ was measured between the transverse contacts in magnetic fields up to 9 T, applied perpendicularly to the sample surface and flow of the applied current density $j$ = 0.5 mA. $V_H(H)$ was calculated according to:
\begin{equation}
V_H(H) = \frac{1}{2}[V(H) - V(-H)],
\end{equation}
and the Hall resistivity was calculated from
\begin{equation}
\rho_H(H,T) = \frac{V_H(H,T)t}{j},
\end{equation}
where \emph{t} = 115 $\mu$m is the thickness of the sample. The obtained isothermal Hall resistivity $\rho_H$ of \NpPtGa~at selected temperatures as a function of magnetic field is presented in Fig. 10 a and b. We observe that in the magnetic field range up to 4 T, $\rho_H$ exhibits a strong field dependence on cooling through the magnetic phase transition. Apparently, below $T_C$ (Fig. 10 a) there appears a S-shaped $\rho_H$, thus  a pronounced maximum is observed in the Hall coefficient,  $R_H(H) = \rho(H)/\mu_0H$ (Fig. 10 c). The $R_H$ maximum shifts down to lower temperatures with decreasing temperature. In the paramagnetic state, the Hall resistivity (Fig. 10 b) shows non-linear field dependence, and steady increases with decreasing temperature.

\par From the isothermal Hall data we have inferred the Hall coefficient in fields of 2 and 9 T as solid symbols in Fig. 11. The data in a field of 9 T agrees well with the 9T-isofield data. A comparison of the 2T- Hall resistivity to the isothermal magnetization data (see Fig. 2 -left panel) suggests that $R_H(T)$ qualitatively behaves like \emph{M(H)}. This behavior is consistent with the view that spin-orbit coupling occurs during resonant scattering at the sites of the Np ions.  Phenomenologically, the Hall coefficient is given by $R_H(H,T) = R_0 + R_s M/\mu_0 H$, where $R_0$ is the ordinary Hall coefficient, which is mainly associated with the Lorentz force on conduction electrons and $R_s$ is the anomalous Hall coefficient, which arises from the side-jump and/or skew scattering.[41,42] Taking into account the Hall data and the isothermal magnetization we obtained values of $R_0$ = -1.98$\times$10$^{-9}$ m$^3/C$ and $R_s$ = 10.33 $\times$10$^{-7}$ m$^3/C$. The fitted parameters reproduce very well the Hall coefficient in the paramagnetic region. Assuming only one type of carriers, $R_0 = 1/|ne|$, the effective carrier density \emph{n} = 3.15$\times$10$^{27}$ m$^{-3}$ can be estimated.
\par In the ferromagnetic range, the Hall coefficients $R_0$ and $R_s$ are temperature dependent. Therefore, in order to separate $R_0$ and $R_s$ from the total $R_H$ we combined isothermal \emph{M} and $R_H$ data (Figs. 3 and 10). We applied a linear regression of the data $\rho_H(H)$ vs. \emph{M(H)}, shown in Fig. 12 a. The resulting $R_0$ and $R_s$ values are plotted as a function of temperature in Fig. 12 b. Typically in f-electron materials the anomalous Hall coefficient is much larger than the ordinary one.[43-45] In the case of \NpPtGa, $R_s$/ $R_0$ reaches a value of $\sim$ 500. It is noticed that both $R_0$ and $R_s$ may have a rapid change below 5 K. At 1.8 K  $R_0$ amounts to 7.69 $\times$ 10$^{-9}$ m$^3$/C, corresponding to one-band carrier concentration of 8.12 $\times$ 10$^{26}$/m$^{-3}$ (or 0.05 carrier/f.u). The small \emph{n} value indicates that \NpPtGa~is a low-carrier system. From the carrier concentration at 1.8 K and the Sommerfeld coefficient of 180 mJ/(mol at. Np K$^2$) we evaluated the effective mass $m^*$ = 83 $m_0$ of the charge carries. We also calculated the Hall mobility $\mu_H =  R_0/\rho$, which appears to vary dramatically with change of temperature. $\mu_H$ falls down by almost two times from 28.3 cm$^2$/Vs at 50 K to 16.8 cm$^2$/Vs  at 1.8 K. Since  $\mu_H \propto \tau/m^*$,  the scattering time $\tau$ of quasi-particles is expected to reduce with lowering temperature.

\section{Discussion}
\par The measured physical properties show that the new intermetallic \NpPtGa~exhibits interesting properties such as the Kondo effect coexisting with ferromagnetic order. First, we discuss the origin of the ferromagnetic interactions in \NpPtGa~in relation to other (U, Np)$_2$TGa$_3$ compounds. We mention that our previous studies have revealed the antiferromagnetic ordering in U$_2$PdGa$_3$ but ferromagnetic one in Np$_2$PdGa$_3$.[4] The evolution of the magnetic ground state in this family of compounds has been  interpreted based on an indirect RKKY exchange mediated via the conduction electrons. Following the RKKY exchange Hamiltonian $H_{ex} \propto -J_{ex}^2\sum_{i,j}F(R_{i,j})S_iS_j$ for the oscillatory interactions between local spin $S_i$ and $S_j$ separated with the interionic distance $R_{ij}$, where $F(R_{i,j}) = [2k_FR_{i,j}{\rm cos}(2k_FR_{i,j}) - {\rm sin}(2k_FR_{i,j})]/R_{ij}^4$, we have generated the RKKY exchange in U$_2$PdGa$_3$ and Np$_2$PdGa$_3$ (dashed lines) in Fig. 13. We used the Fermi wave vector $k_F$ values inferred from the Hall effect data ($k_F$ = 0.93 \AA$^{-1}$ for \UPdGa~and  0.44 \AA$^{-1}$ for \NpPdGa) and assuming $J_{ex} \propto M_{ord}$ ($M_{ord}$ = 0.4 $\mu_B/{\rm U}$ in \UPdGa~and  0.8 $\mu_B/{\rm Np}$ in \NpPdGa~from the magnetization data). In the same manner, we may plot the RKKY exchange energy (solid line in Fig. 13) for \NpPtGa~adopting $k_F$ = 0.29 \AA$^{-1}$ and $M_{ord} \sim$ 0.37$\mu_B/{\rm Np}$.
Considering the range of the interionic distance $R_{ij}$, where two most important distances between magnetic ions in the 213-type compounds, i.e. the nearest neighbor $d_1$ and next-nearest distance $d_2$ take place, we see that the RKKY exchange energy is positive for the Np-based compounds, whereas negative for the uranium one. Thus, the long-range ferromagnetic ordering observed in the Np$_2$TGa$_3$ compounds accords with positive values of $H_{ex}$ and this observation suggests an intimate relationship between the magnetic phase transition temperatures $T_C$ and amplitude of $H_{ex}$ in actinide ferromagnetic Kondo lattices.  Nonetheless, as considered in the INTRODUCTION, these materials have commonly shown high Curie temperatures, in the range 50 - 100 K.  Surprisingly, $T_C$ of \NpPtGa~is found to amount to 26 K only, thus $T_C$ value is decidedly diminished. Therefore, it appears that high Curie temperature may be not general rule for the Kondo actinide ferromagnets. In addition to the reason of partially screening of the S = 1 spins, an aspect on the relative strengths of ferromagnetic interaction and Kondo effect is worthwhile to consider. The view of \NpPtGa~is that $T_K$ is comparable to $T_C$, and on contrary to \NpPtGa, where $T_C$ = 63 K is sizeably greater than $T_K$ = 35 K.

\par We wish to emphasize that the study of specific heat, electrical resistivity and Hall effect of \NpPtGa~in this work and of \NpPdGa~in ref. [4] give evidence that both \NpPdGa~and \NpPtGa~belong to the same family of compounds where the Kondo effect and a ferromagnetism coexist.  A large effective mass of charge carriers with low carrier concentration found in these ferromagnetic Kondo lattices can be understood in terms of the underscreened Kondo-lattice model, developed by Coqblin and coworkers.[19,20] for S = 1 interacting with a spin density of conduction electrons via an on-site antiferromagnetic Kondo coupling. A characteristic parameter for a given Kondo system is its Kondo temperature $T_K$. We estimated the Kondo temperature  $T_K$ = 35 K for \NpPdGa~and 24 K for \NpPtGa. This result is different from what could be expected from the "chemical pressure" effect alone, since the unit cell volume of \NpPdGa~is larger than that of \NpPtGa. In a Kondo lattice model, the Kondo temperature is given by:[46]
 \begin{equation}
 T_K \propto {\rm exp}[-1/|J_{cf}N(E_F)|],
 \end{equation}
where  $N(E_F)$ is  the density of states at the Fermi  level  $E_F$ and the exchange integral $J_{cf}$ is obtained from the  Schrieffer-Wolff transformation of the Anderson model as:[47]
\begin{equation}
J_{cf} \propto \frac{<V_{cf}^2>U}{\varepsilon_f(\varepsilon_f+U)},
\end{equation}
where $<V_{cf}>$ is the matrix element characterizing the hybridization between conduction and \emph{f}  electrons at $E_F$, \emph{U} is the Coulomb integral and $\varepsilon_f$ is the position of the \emph{f} level relative to the Fermi level.  The lowering $T_K$ in Np$_2$PtGa$_3$ compared to Np$_2$PdGa$_3$ would be due a decreasing $J_{cf}$, which relates with an abating f-ligand hybridization or raising the distance between f and $E_F$ levels. The reduction of $J_{cf}$ is articulated when assuming that $T_C \propto  J_{cf}^2$, expected for RKKY interactions.

\section{Conclusions}
\par We have synthesized a novel intermetallic \NpPtGa, which adopts the orthorhombic CeCu$_2$-type crystal structure. The magnetization, electron transport and specific heat data establish that the compound is a ferromagnetic Kondo system with the Curie temperature $T_C \sim$  26 K and the Kondo temperature $T_K \sim$ 24 K. Thus, \NpPtGa~becomes a new member of a relatively rare family of compounds where the Kondo effect and a ferromagnetism coexist. For \NpPtGa~and other actinide-based 213-type compounds, the magnetic ground state would be interpreted qualitatively with the help of the RKKY exchange energy, which de facto is more sensitive to change of $k_F$. It may be remarked that \NpPdGa~and \NpPtGa~both exhibit a large effective mass of the charge carriers and low carrier concentration, and both might be classified to the same family of the ferromagnetic Kondo lattices.  However, the decrease of $T_K$ in \NpPtGa~compared to that in \NpPdGa~would not be interpreted by "chemical pressure" effect. The reason of this is presumably due to a decrease in the exchange between the localized 5f and conduction electrons in the studied compound.

\section{Acknowledgments}
We acknowledge the access to the infrastructure and financial support provided by the European Commission
in the frame of the "Actinide User Laboratory" program at JRC-ITU.\\

\section{References}
$[1]$~V. Sechovsky and L. Havela, Handbook of Magnetic Materials, volume 11, , Ed. K. H. J. Buschow, (North-Holland, Amsterdam, 1998), p. 1.\\
$[2]$~V. H. Tran, J. Phys.: Condens. Matter\textbf{ 8}, 6267 (1996).\\
$[3]$~V. H. Tran, F. Steglich, G. Andre, Phys. Rev. B \textbf{65}, 134401 (2002).\\
$[4]$~V. H. Tran, J.-C. Griveau,  R. Eloirdi,  W. Miiller, and E. Colineau, Phys. Rev. B  \textbf{82}, 094407 (2010).\\
$[5]$~S. K. Malik and D. T. Adroja, Phys. Rev. B \textbf{43}, 6295 (1991).\\
$[6]$~L. Menon and S. K. Malik, Phys. Rev. B \textbf{52}, 35 (1995).\\
$[7]$~E. Bauer, R. Hauser, E. Gratz, G. Schaudy, M. Rotter, A. Lindbaum, D. Gignoux and  D. Schmitt, Z. Phys. B \textbf{92}, 411 (1993).\\
$[8]$~J. Larrea J., M. B. Fontes, A. D. Alvarenga, E. M. Baggio-Saitovitch, T. Burghardt, A. Eichler and M. A. Continentino, Phys. Rev B \textbf{72}, 035129 (2005).\\
$[9]$~T. Takeuchi, A. Thamizhavel, T. Okubo, M. Yamada, N. Nakamura, T. Yamamoto, Y. Inada, K. Sugiyama, A Galatanu, E. Yamamoto, K. Kindo, T. Ebihara, and Y. \={O}nuki, Phys. Rev. B \textbf{67}, 064403 (2003).\\
$[10]$~M. B. Fontes, S. L. Bud$^\prime$ko, M. A. Continentino, E. M. Baggio-Saitovitch, Physica B \textbf{270}, 255 (1999).\\
$[11]$~C. Krellner, N. S. Kini, E. M. Br\"{u}ning, K. Koch, H. Rosner, M. Nicklas, M. Baenitz, and C. Geibel, Phys. Rev B \textbf{76}, 104418 (2007).\\
$[12]$~J. Schoenes, B. Frick and O. Vogt, Phys. Rev. B \textbf{30}, 6578 (1984).\\
$[13]$~Z. Bukowski, R. Tro\'{c}, J. St\c{e}pie\'{n}-Damm, C. Su\l kowski and V.H. Tran, J. Alloys Compd., \textbf{403}, 65 (2005).\\
$[14]$~V. H. Tran, R. Tro\'{c}, Z. Bukowski, D. Badurski and C. Su\l kowski, Phys. Rev. B \textbf{71}, 094428 (2005).\\
$[15]$~D. Kaczorowski, Solid State Commun. \textbf{99}, 949 (1998).\\
$[16]$~J. Schoenes, R. L.  Withers and F. Hulliger, J. Magn. Magn. Mater., \textbf{310}, 1778 (2007).\\
$[17]$~E. Colineau, F. Wastin, J. P. Sanchez, and J. Rebizant, J. Phys.: Condens. Matter  \textbf{20}, 075207 (2008).\\
$[18]$~{Doniach} S. Doniach, Physica B - C\textbf{ 91}, 231 (1977).\\
$[19]$~N.  B.  Perkins,  M.  D.  N\'{u}\~{n}ez-Regueiro,  B.  Coqblin,  and  J.  R. Iglesias, Phys. Rev. B  \textbf{76}, 125101 (2007).\\
$[20]$~C. Thomas, A. S. R. Sim\~{o}es, J. P. Iglesias, C. Lacroix, N. B. Perkins and B. Coqblin, Phys. Rev. B \textbf{83}, 014415  (2011).\\
$[21]$~C. Thomas, A. S. R. Sim\~{o}es, J. P. Iglesias, C. Lacroix and B. Coqblin, J. of Physics: Conf. Ser. \textbf{391}, 012174 (2012).\\
$[22]$~T. Roisnel and J. Rodriguez-Carvjal, in: Materials Science Forum, Proc. Seventh European Powder Diffraction Conference (EPDIC 7), Ed. R. Delhez and E.J. Mittenmeijer, (Barcelona, 2000), p. 118.\\
$[23]$~R. Caciuffo, J. A. Paix\~{a}o, C. Detlefs, M. J. Longfield, P. Santini, N. Bernhoeft, J. Rebizant, and G. H. Lander, J. Phys.: Condens. Matter \textbf{ 15}, S2287 (2003).\\
$[24]$~K. Binder,  A. P. Young,  Rev. Mod. Phys., \textbf{58}, 801 (1986).\\
$[25]$~J. A. Mydosh, Spin Glasses: An experimental Introduction (Taylor \& Francis, London, 1993).\\
$[26]$~V. H. Tran, R. Tro\'{c}, P. de V. du Plessis, G. Andr\'{e} and F. Bour\'{e}e, Phys. Rev. B \textbf{56}, 11065 (1997).\\
$[27]$~J. B. Goodenough, R. I. Dass, J. Zhou, Solid State Sci. \textbf{4}, 297 (2007).\\
$[28]$~G. Liang, F. Yen, S. Keith and M. Croft, J. Magn. Magn. Mater.  \textbf{314}, 52 (2007).\\
$[29]$~S.  Chikazumi,  Physics  of  Ferromagnetism (Oxford  University Press, New York, 1964), p. 274.\\
$[30]$~J. W. Ross and D. J. Lam,  J. Appl. Phys. \textbf{38}, 1451 (1967).\\
$[31]$~P. Erd\"{o}s, G. Solt, Z. \.{Z}o\l nierek, A. Blaise and J. M. Fournier, Physica B+C \textbf{102}, 164 (1980).\\
$[32]$~P. Santini, S. Carretta, N. Magnani, G. Amoretti, and R. Caciuffo, Phys. Rev. Lett. \textbf{97}, 207203 (2006).\\
$[33]$~A. Freeman and  R. E. Watson, Magnetism  Vol. IIa,  ed.  G. T.  Rado  and  H. Suhl  (New York Academic, 1965),
p. 168.\\
$[34]$~D. Aoki, Y. Haga, T. D. Matsuda, N. Tateiwa, S. Ikeda, Y. Homma, H. Sakai, Y. Shiokawa, E. Yamamoto, A. Nakamura, R. Settai, and Y. \"{O}nuki, J. Phys. Soc. Jpn,  \textbf{76}, 063701 (2007).\\
$[35]$~J.-C. Griveau, K. Gofryk, and J. Rebizant, Phys. Rev. B \textbf{77}, 212502 (2008).\\
$[36]$~D. W. Osborne and E. F. Westrum, J. Chem. Phys. \textbf{21}, 1884 (1953).\\
$[37]$~N. Magnani, P. Santini, G. Amoretti, R. Caciuffo, , P. Javorsk\'{y}, F. Wastin, J. Rebizant, G.H. Lander, Physica B \textbf{359 - 361},  1087 (2005).\\
$[38]$~J. C. Bonner and M. E. Fisher, Phys. Rev. \textbf{135}, A640 (1964).\\
$[39]$~V. T. Rajan, Phys. Rev. Lett. \textbf{51}, 308 (1983).\\
$[40]$~B. Cornut and B. Coqblin, Phys. Rev. B\textbf{ 5}, 4541 (1972).\\
$[41]$~J. Smit, Physica (Amsterdam) \textbf{ 24}, 39 (1958).\\
$[42]$~L. Berger, Phys. Rev. B  \textbf{2}, 4559 (1970).\\
$[43]$~J. Schoenes and J. J. M. Franse, Phys. Rev. B \textbf{33}, 5138 (1986).\\
$[44]$~D. Kaczorowski and J. Schoenes, Solid State Commun. \textbf{74}, 143 (1990).\\
$[45]$~V. H. Tran, S. Paschen, R. Tro\'{c}, M. Baenitz, and F. Steglich, Phys. Rev. B \textbf{69}, 195314 (2004).\\
$[46]$~M. Lavagna, C. Lacroix and M. Cyrot,  J. Appl. Phys. \textbf{53}, 2055 (1982).\\
$[47]$~J. P. Schrieffer and P. A. Wolff,  Phys. Rev. \textbf{149}, 491 (1966).\\


\begin{figure}[h]
\includegraphics[scale=0.45]{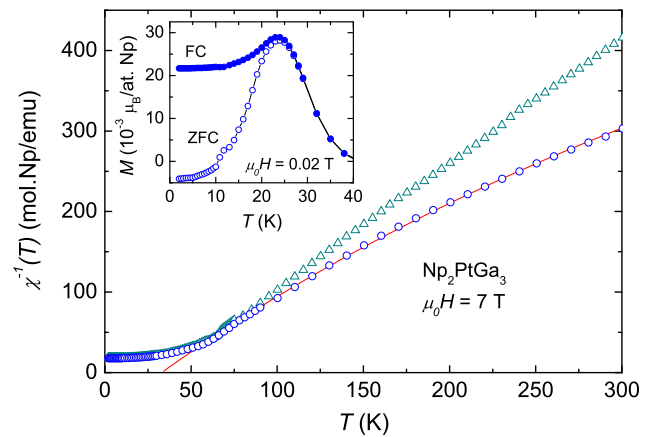}
\caption{\label{1} (color online) Temperature dependence of the inverse magnetic susceptibility of \NpPtGa~measured in a field of 7 T. The solid line is a fit to the modified Curie-Weiss law. The triangle symbols are the data after subtraction of the temperature independent susceptibility $\chi_0$. The inset shows the temperature dependence of the low-temperature magnetization $M(T)$ at 0.02 T, measured in ZFC and FC modes.}
\end{figure}
\begin{figure}[h]
\includegraphics[scale=0.45]{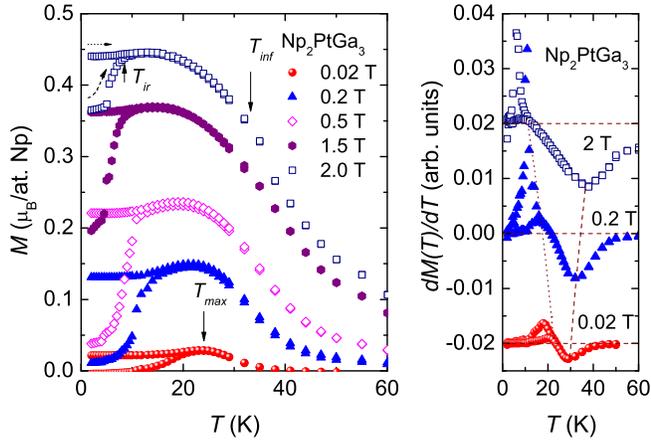}
\caption{\label{2} (color online) Left panel: Temperature dependence of the iso-field magnetization measured in selective fields up to 2 T. The dashed and dotted arrows indicate the data obtained in ZFC and FC modes, respectively. The solid arrows denote the positions of characteristic temperatures $T_{ir}$, $T_{max}$ and $T_{inf}$. Right panel: The temperature derivative of the magnetization \emph{dM(T)/dT} vs temperature. For clarity of the figure, the data at 0.02 and 2 T are shifted vertically by -0.02 and 0.02 $\mu_B$/(at. Np K), respectively. The dotted and dashed lines, as guide to the eye, show the field dependence of $T_{ir}$ and $T_{inf}$, respectively.}
\end{figure}
\begin{figure}[h]
\includegraphics[scale=0.45]{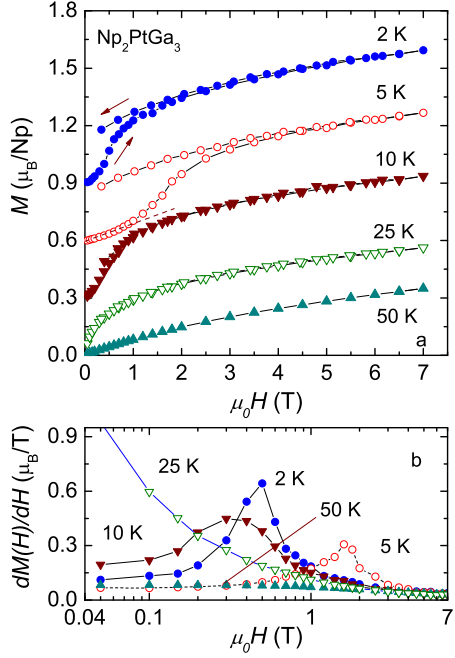}
\caption{\label{3} (color online) a) Isothermal magnetization at various temperatures as a function of applied magnetic fields. For the sake of clarity, the data at 2, 5 and 10 K are shifted upwards by 1.8, 1.2 and 0.6 $\mu_B$, respectively. The dotted line denotes a linear field dependence of the initial magnetization at 2 K. b) Field derivative of \emph{M(H)} vs applied magnetic fields.}
\end{figure}
\begin{figure}[h]
\includegraphics[scale=0.45]{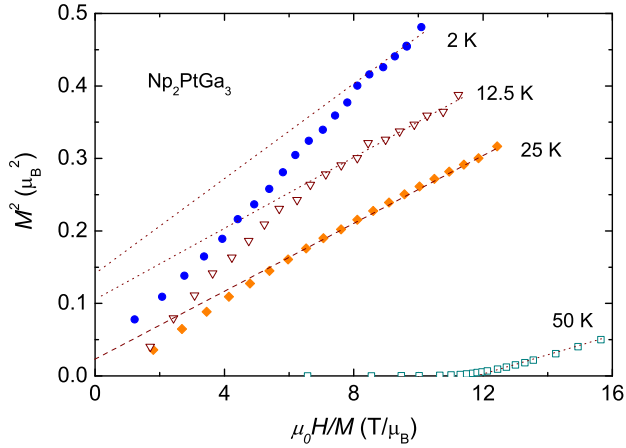}
\caption{\label{4} (color online) Arrott plot for selected magnetization data obtained during decreasing applied magnetic fields. The dotted and dashed lines are a guide for eyes.}
\end{figure}
\begin{figure}[h]
\includegraphics[scale=0.45]{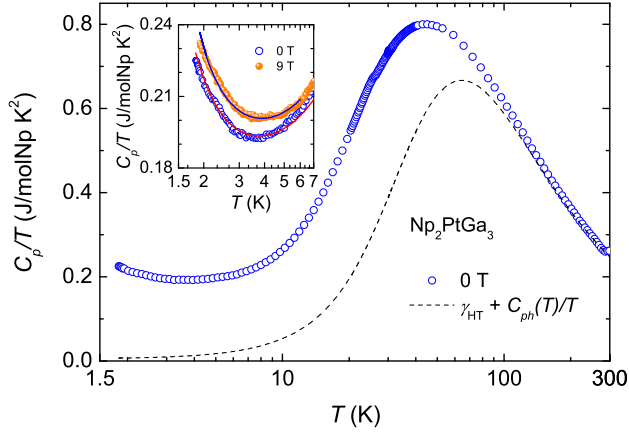}
\caption{\label{5} (color online)
Temperature dependence of the specific heat divided by temperature of \NpPtGa~at $\mu_0H$ = 0 T. The dashed line represents the sum of the phononic and high-temperature electronic contributions. The inset shows the low-temperature data obtained in 0 and 9 T with the results of fitting of the nuclear, phononic and electronic contributions. }
\end{figure}
\begin{figure}[h]
\includegraphics[scale=0.45]{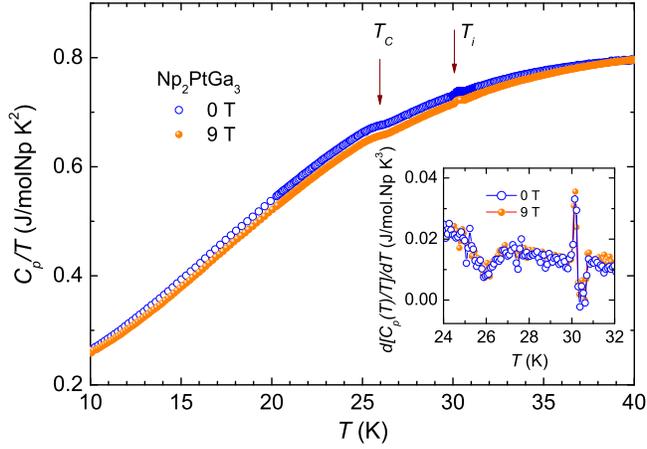}
\caption{\label{6} (color online)
Temperature dependence of the specific heat divided by temperature of \NpPtGa~in $\mu_0H$ = 0 and 9 T.  The inset shows the temperature derivative of $C_p/T$ in 0 and 9 T.}
\end{figure}
\begin{figure}[h]
\includegraphics[scale=0.45]{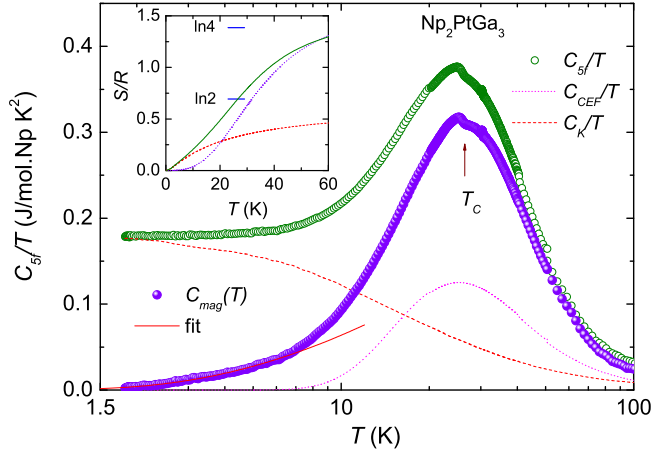}
\caption{\label{7} (color online)
Temperature dependence of the 5f-electron  ( open symbols) and magnetic ( closed symbols) specific heats divided by temperature of \NpPtGa. The dashed line is the Kondo specific heat with the Kondo temperature $T_K$ = 24 K. The dotted line is the specific heat due crystal electric field splitting with an energy $\Delta_{CEF}/k_B$ = 82 K. The inset shows the 5f-electron (solid line), magnetic (dotted line) and Kondo (dashed line) entropies as a function of temperature.}
\end{figure}

\begin{figure}[h]
\includegraphics[scale=0.45]{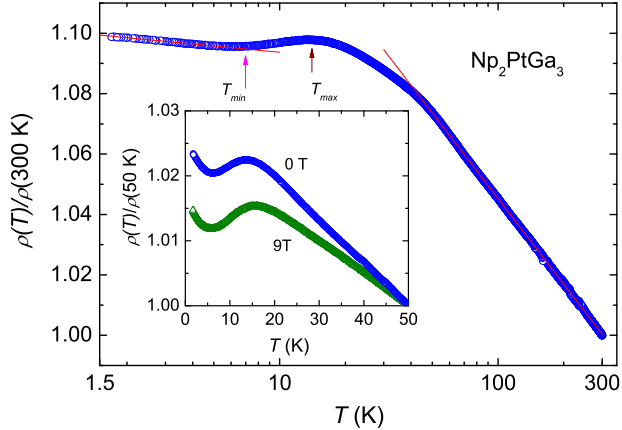}
\caption{\label{8} (color online)
Temperature dependence of the normalized electrical resistivity $\rho(T)/\rho(300 {\rm K})$ of \NpPtGa~in zero field. The solid lines are fits with -ln\emph{T} dependencies. The inset shows field dependencies of the normalized electrical resistivity $\rho(T)/\rho(50 {\rm K})$.}
\end{figure}

\begin{figure}[h]
\includegraphics[scale=0.45]{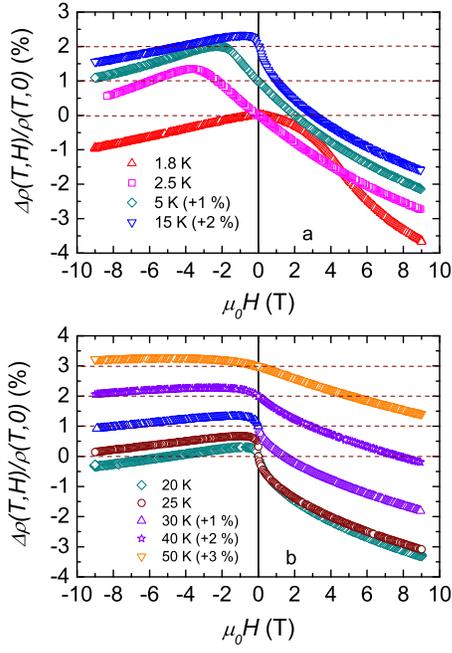}
\caption{\label{9} (color online)
Magnetoresistance as a function of the magnetic field at different temperatures. The data at 5 and 15 K are shifted vertically by 1 and 2 $\%$, and data at 30, 40 and 50 K by 1,  2 and 3 $\%$, respectively.}
\end{figure}

\begin{figure}[h]
\includegraphics[scale=0.45]{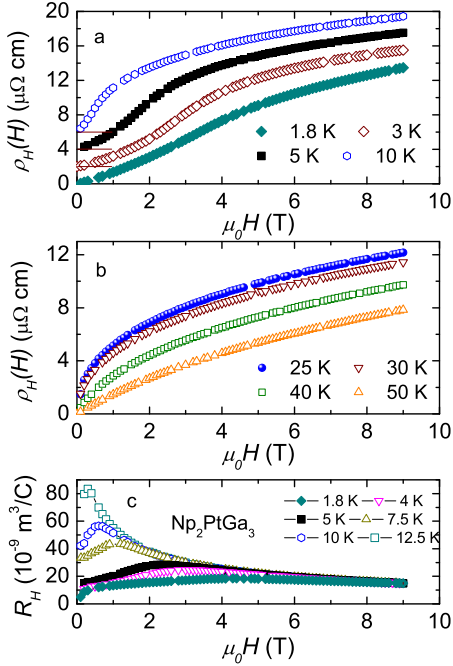}
\caption{\label{10} (color online)
a and b) Hall resistivity of \NpPtGa~ as a function of the magnetic field at different temperatures. The data at 3, 5 and and 10 K are shifted vertically by 2, 4 and 6 $\mu \Omega$ cm, respectively. c)  The field dependence of the Hall coefficient of \NpPtGa~in the ordered state.}
\end{figure}
\begin{figure}[h]
\includegraphics[scale=0.45]{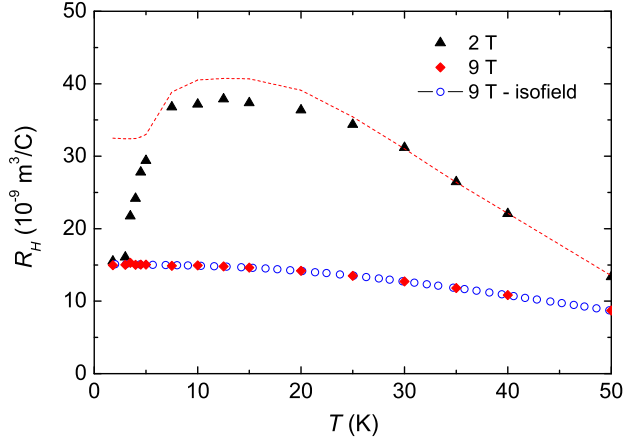}
\caption{\label{11} (color online)
Temperature dependence of the isothermal (closed symbols) in fields of 2 and 9 T, and 9T-isofield (open circles) Hall coefficients. The dashed line is a fit to the isothermal 2T data.}
\end{figure}

\begin{figure}[h]
\includegraphics[scale=0.45]{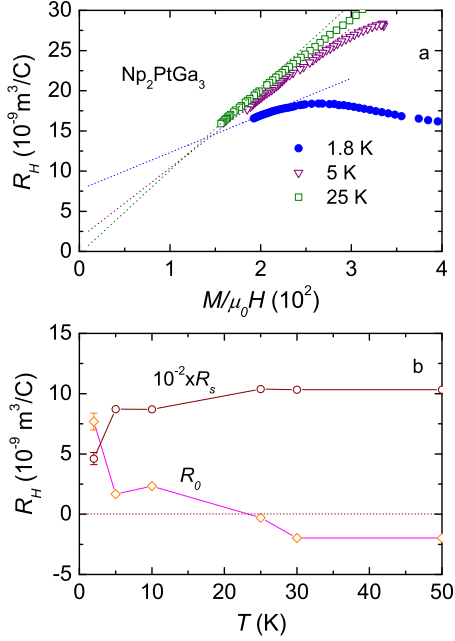}
\caption{\label{12} (color online)
a) The Hall coefficient as a function of $M/\mu_0H$. The dashed lines are linear fits to 1.8 K and 50 K data. b) The temperature dependencies of the ordinary $R_0$, anomalous $R_s$ and experimental $R_H$ (solid squares) Hall coefficients of \NpPtGa. The dashed line is the theoretical line, simulated from the experimental $M/\mu_0H$ and fitted $R_0$, $R_s$ values.}
\end{figure}
 \begin{figure}[h]
\includegraphics[scale=0.5]{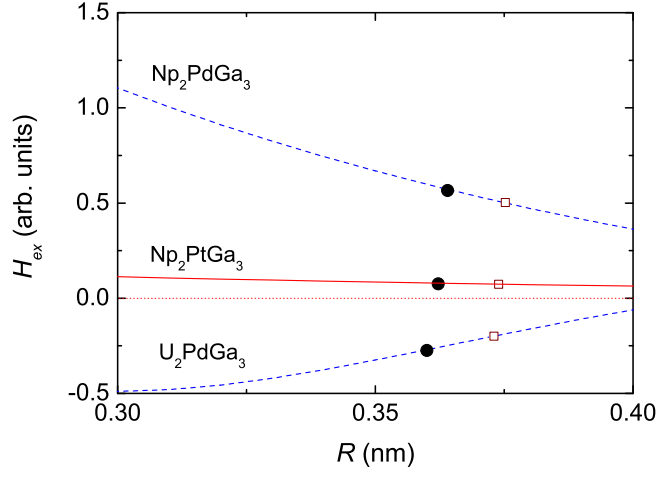}
\caption{(color online) The RKKY exchange energy $H_{ex}$ as a function of distance $R_{i,j}$ in \UPdGa, \NpPdGa~and \NpPtGa. The exchange between nearest, next-nearest magnetic ions are indicated by closed circles and squares, respectively. }
\label{13}
\end{figure}

\bibliography{}
\end{document}